\documentclass{article}

\usepackage{arxiv}

\usepackage[utf8]{inputenc} % allow utf-8 input
\usepackage[T1]{fontenc}    % use 8-bit T1 fonts
\usepackage{hyperref}       % hyperlinks
\usepackage{url}            % simple URL typesetting
\usepackage{booktabs}       % professional-quality tables
\usepackage{amsfonts}       % blackboard math symbols
\usepackage{nicefrac}       % compact symbols for 1/2, etc.
\usepackage{microtype}      % microtypography
\usepackage{amsmath}
\usepackage{cleveref}       % smart cross-referencing
\usepackage{tabularx}
\usepackage{lipsum}         % Can be removed after putting your text content
\usepackage{graphicx}
\usepackage{natbib}
\usepackage{doi}
\usepackage{amssymb}  
\usepackage{pifont}   % 提供 \ding

\title{Learning Economic Model Predictive Control via Clustering and Kernel-Based Lipschitz Regression}

\usepackage{authblk}

\newbox{\orcid}\sbox{\orcid}{\includegraphics[scale=0.06]{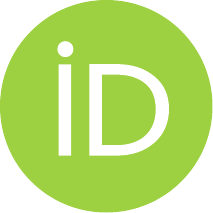}} 
\author[1]{%
	\href{https://orcid.org/0000-0002-2476-8941}{\usebox{\orcid}\hspace{1mm} Weiliang Xiong }%
}
\author[1]{%
	Defeng He \thanks{\texttt{Corresponding Author: Defeng He: hdfzj@zjut.edu.cn; This work was supported by the National Natural Science Foundation of China (62173303, U24A20270).
}}}%
\author[2]{%
	Haiping Du }%
\affil[1]{College of Information Engineering, Zhejiang University of Technology, Hangzhou, China 310023}
\affil[2]{School of Electrical, Computer and Telecommunications Engineering, University of Wollongong, Wollongong, Australia, 2522}

\renewcommand{\shorttitle}{\textit{arXiv} Template}

\hypersetup{
pdftitle={Learning Economic Model Predictive Control via Clustering and Kernel-Based Lipschitz Regression},
pdfsubject={q-bio.NC, q-bio.QM},
pdfauthor={Weiliang Xiong},
pdfkeywords={First keyword, Second keyword, More},
}

\begin{document}
\maketitle

\begin{abstract}
	This paper presents a novel learning economic model predictive control scheme for uncertain nonlinear systems subject to input and state constraints and unknown dynamics. We design a fast and accurate Lipschitz regression method using input and output data that combines clustering and kernel regression to learn the unknown dynamics. In each cluster, the parallel convex optimization problems are solved to estimate the kernel weights and reduce the Lipschitz constant of the predictor, hence limiting the error propagation in the prediction horizon. We derive the two different bounds of learning errors in deterministic and probabilistic forms and customize a new robust constraint-tightening strategy for the discontinuous predictor. Then, the learning economic model predictive control algorithm is formulated by introducing a stabilized optimization problem to construct a Lyapunov function. Sufficient conditions are derived to ensure the recursive feasibility and input-to-state stability of the closed-loop system. The effectiveness of the proposed algorithm is verified by simulations of a numerical example and a continuously stirred tank reactor. 
\end{abstract}

% keywords can be removed
\keywords{Nonlinear systems \and Predictive control  \and Robust control \and Learning control \and Stability}

\section{Introduction}
Economic Model Predictive Control (EMPC) has gained significant attention in various applications, e.g., energy systems \citep{clarke2018hierarchical, wu2022economic}, transportation systems \citep{farooqi2020shrinking, luo2022multiobjective}, and chemical processes \citep{ellis2014tutorial,ellis2017economic}, owing to its ability to handle constraints and nonlinearities while optimizing economic performance explicitly. The economic performance metric is not necessarily positive definite related to any steady state. Therefore, the closed-loop system under EMPC often presents state divergence or periodic oscillations if the economic cost is optimized directly \citep{ellis2014tutorial,ellis2017economic}. 

Multiple EMPC formulations are developed to address the stability issue. For example, Diehl et al. \citep{diehl2010lyapunov} adopted the strong duality assumption to design stabilizing EMPC, which was extended to the dissipativity methods in \citet{angeli10average,gros2022economic,faulwasser2018economic}. Using the controllability condition and a long enough horizon, the closed-loop system with EMPC converges toward a neighbourhood of the optimal steady state \citep{grune2013economic,grune2014asymptotic}. Without the dissipativity assumption, \citet{maree2016combined} weighted the stabilized and economic criteria to ensure the stability of the resulted system with EMPC. \citet{alamir2021new} utilized weighted state increments to achieve convergence toward the steady-state manifold. In \citet{faulwasser2018economic}, the stability constraint of a prior Lyapunov function rendered the asymptotical stability. To avoid the artificial selection of the weights or obtaining prior Lyapunov functions, \citet{he2015stability, he2016economic} proposed stabilizing EMPC schemes with contractive constraints of Lyapunov-like functions, which has been extended to robust EMPC and event-triggered EMPC in \citet{defeng2019input, he2023event}. These results, however, assume that the exact dynamic is known to construct the Finite Horizon Optimal Control Problem (FHOCP) of EMPC. Unfortunately, the first principle-based modelling is challenging in actual applications and usually induces approximate models with considerable uncertainties.

Due to the ability to accurately identify complex nonlinear mappings using big data, several machine learning algorithms are used to construct predictive models \citep{xiong2023adaptive,zhang2022robust, murphy2022probabilistic, aswani2013provably, bongard2022robust, hewing2019cautious, calliess2020lazily, calliess2018nonlinear, carnerero2023kernel} (i.e., predictors) from industrial data, such as Koopman operator \citep{zhang2022robust, mei2024input}, Fundamental Lemma \citep{bongard2022robust} and Nadaraya-Watson estimator  \citep{aswani2013provably}. However, in these schemes, deterministic error bounds are not derived for robust theory analysis, and thus Input-to-State Stability (ISS) cannot be strictly guaranteed. To this end, \citet{calliess2014conservative} proposed a specialized nonparametric regression approach named Lipschitz interpolation \citep{ calliess2020lazily, calliess2018nonlinear}, which can indirectly provide learning error bounds by ensuring the Lipschitz continuity of predictors. The Lipschitz interpolation was generalized to the Kinky Inference (KI) by using Hölder continuity in \citep{calliess2014conservative}. However, the KI-class methods encounter the representative trade-off problem of computational burden and prediction accuracy as improving accuracy entails exponential increase of the volume of data, which aggravates the calculation burden of online prediction \citep{calliess2014conservative, manzano2019output, manzano2021componentwise}.

The computation burden of KI consists of offline optimization of Lipschitz parameters and online prediction \cite{calliess2014conservative, manzano2019output,manzano2021componentwise}. For offline optimization, Calliiess \cite{calliess2017lipschitz} used Lipschitz optimization, and Calliiess et al. \cite{calliess2020lazily} proposed the lazily iterative calculation method with quadratic complexity to reckon Lipschitz constants. Nevertheless, the computation complexity still remarkably increases with growing data and then limits the scalability of big data. For online prediction, the complexity of nonparametric KI increases linearly with the volume of data. To alleviate the burden, Manzano et al. \cite{manzano2019output, blaas2019localised} offline divided data into hyperrectangulars and KI was used in each data subset to reckon local predictions. 

On the other hand, the prediction accuracy of predictors must satisfy the requirements of robust MPC to guarantee feasibility and ISS. For improving accuracy, Manzano et al. \cite{manzano2021componentwise} employed the componentwise Hölder KI to refine system output bound estimations, with the challenge of calculating optimal Hölder parameter matrices. Maddalena and Jones \cite{maddalena2020learning} proposed the Smooth Lipschitz Regression (SLR) method to improve the accuracy of predictors by using kernel technique. However, the dimension of the kernel optimization problem is at least linearly related to the data amount, with computational intractability in big data. Therefore, it is desirable to design a new KI learning algorithm that ensures desired properties, such as fast prediction, high accuracy and error boundary.

This paper proposes a novel Learning EMPC (LEMPC) scheme for uncertain nonlinear systems subject to input and state constraints and unknown dynamics. A schematic of the CKLR-based LEMPC is outlined in Fig. 1. In the offline phase, the Clustering and Kernel-based Lipschitz Regression (CKLR) approach is designed to reconstruct an accurate predictor with fast prediction. The clustering method adaptively divides the data and constraint set, and the kernel weights are optimized to derive the predictor with high accuracy and a low Lipschitz constant. Then, in the online phase, the LEMPC scheme is designed by sequentially solving the economic FHOCP and stable FHOCP. Under some assumptions of disturbances, the prediction error bounds of the CKLR-based predictor are explicitly derived. Then a new constraint-tightening method is presented to ensure the robustness of the LEMPC controller. Moreover, some sufficient conditions are derived to ensure the ISS of the uncertain closed-loop system with the LEMPC. The superiority of the LEMPC is verified by a numerical example and a continuously stirred tank reactor (CSTR).

\begin{figure}
	\centering
	\includegraphics[width=.5\textwidth]{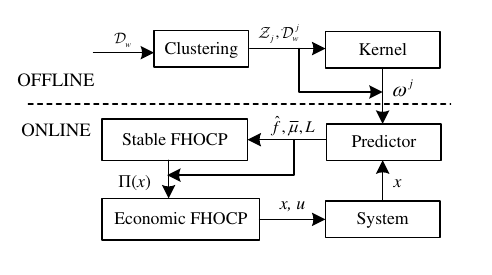}
	\caption{A schematic of the CKLR-based LEMPC scheme }
	\label{FIG:1}
\end{figure}
The main contributions of this work are summarized as follows:

1) A customized CKLR method combining clustering and kernel methods is developed to extend traditional KI-class methods, effectively addressing big data tractability, fast online prediction, and high accuracy. 

2) Some properties of the CKLR predictor are analyzed, e.g., continuity, smoothness, and error bound, to propose a novel constraint-tightening method to guarantee the robustness of the controller.  

3) A new LEMPC algorithm is proposed, incorporating adjusted parameters to balance economic performance and stabilization speed. It is especially tailored for unknown nonlinear systems and includes traditional stabilized MPC as a special case under appropriate parameter design. Moreover, sufficient conditions of feasibility and ISS are derived.

This paper is organized as follows: The problem description and basic definitions are presented in Section 2. The whole learning algorithm, with analysis of continuity, smoothness and error bounds, is developed in Section 3. The LEMPC design with its theoretical property analysis is provided in Section 4. The simulation examples are given in Section 5, and Section 6 concludes this work. 

\noindent
\textit{\textbf{Notation:}} $R^n$ represents the $n$-dimensional real space, ${I_{ \geqslant i}}$ is the set of integers not less than $i$ and $I_a^b: = \{ a,a + 1...,b\} $, The ${x_{k + t}}$ is the actual state of the system at time $k+t$, distinguishing from $t$ time step ahead prediction at time $k$, ${x_{t|k}}$. Define $|x|$ as the absolute value, $||x||$ as 2-norm. For two given sets, $A$ and $B$, the Cartesian product is $A \times B$, Minkowski set addition, and Pontryagin set difference are $A \oplus B = \{ a + b|a \in A,b \in B\} $,
$A \ominus B = \{ a|\{ a\}  \oplus B \subseteq A\}$, respectively. A function $\gamma :{R_{ \geqslant 0}} \to {R_{ \geqslant 0}}$ is a $\mathcal{K}$-function if it is continuous, strictly increasing, and $\gamma (0) = 0$. A $\mathcal{K}$-function $\gamma$ is ${\mathcal{K}_\infty }$-function if it is a $\mathcal{K}$-function and $\gamma (s) \to \infty$ when $s \to \infty $. A function $\gamma :{R_{ \geqslant 0}} \times {I_{ \geqslant 0}} \to {R_{ \geqslant 0}}$ is a $\mathcal{K}\mathcal{L}$-function if, for each $k$, $\gamma (\cdot,k)$ is a $\mathcal{K}$-function, $\gamma (s,\cdot)$ is nonincreasing, and $\gamma (s, 0) \to \infty $ when $s \to \infty $. 

\section{Problem description and preliminary}
Consider an unknown discrete-time nonlinear system 
\begin{equation}
{x_{k + 1}} = f({x_k},{u_k}) + {\delta _k},
\end{equation}
where $k \in {I_{ \geqslant 0}},x \in {R^n},u \in {R^m}$ and $\delta  \in {R^n}$ are the time instant, state variable, control input and additive disturbance. The dynamic $f:{R^n} \times {R^m} \to {R^n}$ is unknown. Assume the system state $x_k$ is measurable for feedback and control. The system (1) is subject to the following constraint: 
\begin{equation}
({x_k},{u_k}) \in \mathcal{Z},\quad \forall k \in {I_{ \geqslant 0}},
\end{equation}
\noindent
where $\mathcal{Z}$  is a compact set and contains the origin in interior.  

In what follows, some definitions and lemmas are recalled. 

\noindent
\textbf{Definition 1.} \citep{jiang2001input} The system ${x_k}_{ + 1} = f({x_k}) + {\delta_k}$ is ISS if there exist ${\mathcal{K}_\infty }$-function ${\alpha _1}$, $\mathcal{K}\mathcal{L}$-function $\beta$, such that 
\begin{equation}
\left\| {{x_k}} \right\| \leqslant \beta (\left\| {{x_0}} \right\|,k) + {\alpha _1}(\bar \delta ),\quad \forall k \in {I_{ \geqslant 0}},
\end{equation}
for $\left\| {{\delta _k}} \right\| \leqslant \bar \delta ,\forall k \in {I_{ \geqslant 0}}$.

\noindent
\textbf{Lemma 1.} \citep{jiang2001input}  The system ${x_{k + 1}} = f({x_k}) + {\delta _k}$ is ISS if there exist ${\mathcal{K}_\infty }$-functions ${\alpha _1},{\alpha _2},{\alpha _3},$  $\mathcal{K}$-function $\rho $ and an ISS-Lyapunov function $V:{R^n} \to R$ such that 
\begin{subequations}
\begin{equation}
{\alpha _1}(\left\| {{x_k}} \right\|) \leqslant V(x) \leqslant {\alpha _2}(\left\| {{x_k}} \right\|),
\end{equation}
\begin{equation}
V({x_{k + 1}}) - V({x_k}) \leqslant  - {\alpha _3}(\left\| {{x_k}} \right\|) + \rho \;(\bar \delta ),
\end{equation}
\end{subequations}
for $\left\| {{\delta _k}} \right\| \le \bar \delta ,\forall k \in {I_{ \ge 0}}$.

\noindent
\textbf{Remark 1.} ISS implies that the system is ultimately bounded by a $\mathcal{K}$-function of the disturbance bound. Specifically, the system is asymptotically stable for disturbances that converge to a constant.

The control problem of this paper is to stabilize the unknown system (1) in the presence of constraint (2) and meanwhile, optimizing the economic cost over the finite horizon  $N>0$, i.e., 
\begin{equation}
{J_e}({x_0},{\mathbf{u}_0}) = \sum\nolimits_{i = 0}^{N - 1} {{L_e}({x_i},{u_i})}.
\end{equation}
where the stage cost ${L_e}:{R^n} \times {R^m} \to R$ may be nonconvex or non-positive definite.

It is worth noting that the minimization of (5) usually leads to some unstable behaviours of the closed-loop system. In particular, the dissipativity of system (1) cannot be verified since the dynamics of (1) are unknown. In this paper, we customize a new CKLR method to learn the unknown dynamics of (1) and design the CKLR-based LEMPC to optimize (5) and guarantee the ISS of the closed-loop system. 

\section{CKLR-based model learning}
Consider a given data set ${\cal D} = \{ {w_i},{y_i}\} ,i \in I_1^{{N_{\cal D}}}$ containing ${N_{\cal D}}$ data, where $w = {{\rm{[}}{x^\mathrm{T}},{u^\mathrm{T}}{\rm{]}}^\mathrm{T}}$ is the learning model input, and $y$ is the noise-corrupted state successor, i.e., the corresponding model output. Such a data set ${\cal D}$ could be obtained by using chirp signals to stimulate the system \citep{manzano2020robust}, beyond the scope of this paper. Subsections 3.1-3.2 describe the clustering and kernel technique details, and subsections 3.3-3.4 explain the computational complexity and analyze the predictor properties, respectively. 

\subsection{\textit{K}-means clustering}
In the context of large amounts of data, it is computationally challenging for traditional KI-class methods to solve the global Lipschitz parameter and achieve fast online predictions \citep{manzano2019output,manzano2021componentwise,blaas2019localised}. Hence, we first partition the given ${\cal D}$ by using \textit{K}-means, owing to its scalability and simplicity. Despite this, the CKLR algorithm allows for advanced clustering techniques as long as the data can be partitioned, and the corresponding clustering interfaces are provided.

Define the input set ${{\cal D}_w} = \{ {w_i}\} ,i \in I_1^{{N_{\cal D}}}$ and its partition ${{\cal D}_w} = \bigcup\nolimits_{j = 1}^K {{\cal D}_w^j} ,{\rm{ }}$ ${\cal D}_w^j = \{ {w_i}\} ,i \in I_1^{N_{\cal D}^j}$. Denote the geometric center of ${\cal D}_w^j$  as $z_j$ for $j \in I_1^K$. Then, for each query input $w_i$, one can find the closest center of the cluster by \citep{cheng2020fuzzy}
\begin{equation}
z_i^* = \arg \mathop {\min }\limits_{{z_j}} {\left\| {{w_i} - {z_j}} \right\|^2},i \in I_1^{{N_{\cal D}}},j \in I_1^K.
\end{equation}
From \citet{murphy2022probabilistic}, the cluster centers are updated by 
\begin{equation}
z_j = \frac{1}{{N_\mathcal{D}^j}}\sum\limits_{i:z_i^* = j}^{} {{w_i}} ,j \in I_1^K.
\end{equation}
If the centers of all clusters do not change or the pre-defined iteration number of clustering is finished, then the clustering operation ends; otherwise, repeat the iterative process (6), (7).
\begin{figure}  % 总体流程图
	\centering
	\includegraphics[width=.5\textwidth]{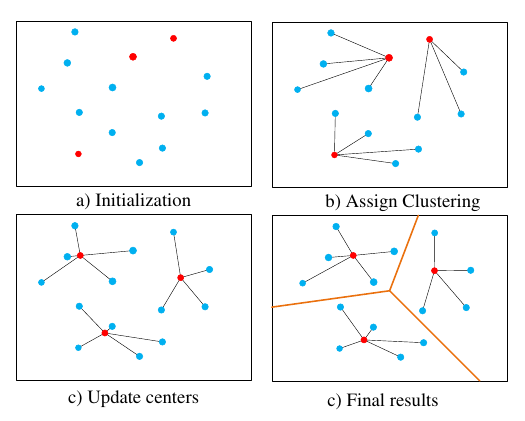}
	\caption{\textit{K}-means clustering, where red points are centers, blue points are data, and the orange lines are interfaces}
	\label{FIG:2}
\end{figure}

Fig.2 shows a visualization of \textit{K}-means clustering. Once it terminates, the constraint set $\mathcal{Z}$ is divided into $K$ subsets, i.e., 
\begin{subequations}
\begin{equation}
\mathcal{Z} = \bigcup\nolimits_{j = 1}^K {{\mathcal{Z}_j}} ,
\end{equation}
\begin{equation}
{\mathcal{Z}_{{j_1}}} \cap {\mathcal{Z}_{{j_2}}} = {\text{ }}\emptyset ,{j_1} \ne {j_2},{j_1} \in I_1^K,{j_2} \in I_1^K.
\end{equation}
\end{subequations}

Note that any tie-breaking method can be used to randomly assign a cluster if a query point falls into the interface. In what follows, we consider the regression in each data subset.
\subsection{Kernel regression}
As the disturbance in the horizon exponentially propagates with the Lipschitz constant of system (1), reducing the learning error and Lipschitz constant of the predictor are desirable. To do this, we introduce the nonlinear kernel to improve accuracy.

Consider the $N_\mathcal{D}^j$  data in $j$th cluster, the local Lipschitz constant of (1) is estimated as 
\begin{equation}
{\bar L^j} = \max \left\{ {0,\mathop {\max }\limits_{1 \leqslant {i_1},{i_2} \leqslant N_\mathcal{D}^j} \frac{{\left\| {y_{{i_1}}^j - y_{{i_2}}^j} \right\| - \lambda }}{{\left\| {w_{{i_1}}^j - w_{{i_2}}^j} \right\|}}} \right\},j \in I_1^K.
\end{equation}
where $i_1$ and $i_2$ are data indexes, the hyper-parameter $\lambda$ is used to compensate for the effect of noise $\delta $ to prevent ${\bar L^j}$ from being infinite. \citet{calliess2020lazily} suggest $\lambda  \geqslant 2\bar \delta $ if $\bar \delta $ is known. Moreover, the predictor $\hat f:{R^{n + m}} \to R$ is chosen as an element of Hilbert space spanned by $N_\mathcal{D}^j$ kernel functions: 
\begin{equation}
\hat f(w) = \sum\nolimits_{i = 1}^{N_\mathcal{D}^j} {\omega _i^j\phi _i^j(w)} ,i \in I_1^{N_\mathcal{D}^j}.
\end{equation}
where $\phi _i^j:{R^{n + m}} \to R$ and $\omega _i^j$ are the kernel function and weight corresponding to data $i$ in cluster $j$, respectively. This paper selects the kernel function as the radial basis function:
\begin{equation}
\phi _i^j(w) = {\sigma ^2}\exp \left( {{{ - {{\left\| {w - w_i^j} \right\|}^2}} \mathord{\left/
 {\vphantom {{ - {{\left\| {w - w_i^j} \right\|}^2}} {2{l^2}}}} \right. \kern-\nulldelimiterspace} {2{l^2}}}} \right),
\end{equation}
where ${\sigma}^2$ and $l$ are the user-designed hyperparameters. Then the optimal ${\omega ^j} = {[\omega _1^j\;\omega _2^j \cdots \omega _{N_\mathcal{D}^j}^j]^{\text{T}}}$ is determined by solving the optimization problem:
\begin{subequations}
\begin{equation}
\mathop {\min }\limits_{{\omega ^j} \in {R^{N_\mathcal{D}^j}}} \;\left\| {{\omega ^j}} \right\|_2^2
\end{equation}
\begin{equation}
{\text{s}}{\text{.t}}{\text{.}}\;{\omega ^j}^T{\phi ^j}({w_i^j}) - y_i^j - {\bar \delta _s} \leqslant 0,
\end{equation}
\begin{equation}
y_i^j - {\bar \delta _s} - {\omega ^j}^\mathrm{T}{\phi ^j}({w_i^j}) \leqslant 0,
\end{equation}
\begin{equation}
\left\| {\sum\nolimits_{i = 1}^{N_\mathcal{D}^j} {{\omega _i}\nabla \phi _i^j({w_s})} } \right\| - {\bar L^j} \leqslant 0,\;\forall s \in I_1^S,
\end{equation}
\end{subequations}
in which the slack factor ${\bar \delta _s}$ is used to ensure the feasibility, the cost (12a) is a convex $L2$ regularization preventing overfitting, the linear inequalities (12b) and (12c) are used to ensure learning accuracy and the quadratic constraints (12d) limit the Lipschitz upper bound of (10) in additional $S$ data points. Therefore, the problem (12) is a convex Quadratically Constrained Quadratic Programming (QCQP) problem \citep{boyd2004convex}. 

Note that the constraint (12d) essentially uses $S$ additional data points to approximate the set constraint 
\begin{equation}
\left\| {\sum\nolimits_{i = 1}^{N_\mathcal{D}^j} {{\omega _i}\nabla \phi _i^j(w)} } \right\| \leqslant {\bar L^j},\;\forall w \in {\mathcal{Z}_j}.
\end{equation}
Theoretically, if the set ${\mathcal{Z}_j}$ can be traversed by infinite $S$ points, then ${\bar L^j}$ is the actual Lipschitz parameter of the predictor. In practice, the Lipschitz parameter is posteriorly estimated if $S$ is finite, which is explained later in Subsection 3.4. 

\noindent
\textbf{Remark 2.} The problem (12) is a feasibility problem if the cost function (12a) is chosen as a constant. Moreover, $L1$ regularization can introduced to realize sparsity and simplify the predictor \citep{boyd2004convex}. Note that the kernel method can approximate any continuous and bounded function as ${N_\mathcal{D}} \to \infty$ \citep{calliess2020lazily}. 

The whole CKLR method is summarized in Algorithm 1.
\begin{center}
\centering
\begin{tabularx}{\linewidth}{X}
\toprule
\textbf{Algorithm 1: CKLR Learning Algorithm (Offline)} \\
\midrule
Step 1. Given the data set ${\cal D}$, set the cluster number $K$, kernel parameters ${\sigma ^{\rm{2}}}$ and $l$, and hyper-parameter $\lambda$.   \\
Step 2. Use the $K$-means method to obtain the clustering centers and data subsets.  \\
Step 3. Set parameter $\lambda$ and calculate the local Lipschitz constant of each output in every clustering. Set parameters $S$ and $\bar \delta _s$ and solve the optimization problem (12) to obtain the optimal weights of the predictor in (10). \\
\bottomrule
\end{tabularx}
\end{center}

From Algorithm 1, once the \textit{K}-means is completed and corresponding weights are obtained, the CKLR predictor can calculate the output of the query point by finding the nearest cluster center $z_j$,$j \in I_1^K$ and calculate the kernel function (10).

\subsection{Computational complexity analysis}
This subsection evaluates the computational complexity of CKLR. For simplicity, it is assumed that each cluster has the same amount of data, and one can adjust $K$ to adapt the size of $\mathcal{D}$. 

Let us first consider the offline training operation. The computational steps of this operation are counted by
\begin{equation}
{T_{off}} = KT{N_\mathcal{D}}a + {\left( {{{{N_\mathcal{D}}} \mathord{\left/
 {\vphantom {{{N_\mathcal{D}}} K}} \right.
 \kern-\nulldelimiterspace} K}} \right)^2}Kb + Kc \in \mathcal{O}\left( {K{N_\mathcal{D}} + {{N_\mathcal{D}^2} \mathord{\left/
 {\vphantom {{N_\mathcal{D}^2} K}} \right.
 \kern-\nulldelimiterspace} K}} \right)
\end{equation}
\noindent
where $T$ is the pre-defined iteration number of clustering, the algorithm-dependent factors $a > 0,b > 0,c > 0$ \citep{manzano2019output}, and the three terms to the right of the equation separately correspond to \textit{K}-means \citep{murphy2022probabilistic}, calculating the Lipschitz constant via (9) and solving the optimization problem (12).

The online prediction operation consists of separately determining the cluster and calculating output via (10) and (11). The computational steps of the one online prediction are
\begin{equation}
{T_{on}} = Ke + {{{N_\mathcal{D}}d} \mathord{\left/
 {\vphantom {{{N_\mathcal{D}}d} K}} \right.
 \kern-\nulldelimiterspace} K} \in \mathcal{O}\left( {K + {{{N_\mathcal{D}}} \mathord{\left/
 {\vphantom {{{N_\mathcal{D}}} K}} \right.
 \kern-\nulldelimiterspace} K}} \right),
\end{equation}
with the algorithm-dependent factors $d>0$ and $e>0$.

Here, the computation steps of online prediction in (15) are similar to the Projected KI (PKI) or Smooth PKI (SPKI) \citep{manzano2019output} but are less than that of KI \citep{calliess2014conservative} for a large $K$. The offline computation steps may slightly increase due to solving the convex problem (12). =We underline that the merits of the CKLR method are its scalability of big data and the improvement in accuracy. However, to design the subsequent LEMPC algorithm with ISS guarantee, the error bounds and actual Lipschitz parameter must be provided explicitly, which is explained in the following subsection.

\subsection{ Properties of predictors }
We first analyze the smoothness and discontinuity of predictor (10).
 
\noindent
\textbf{Lemma 2.} Consider the predictor (10), the Lebesgue measure of set of non-smooth points in $\mathcal{Z}$ is zero. 

\noindent
\textbf{Proof.} The set $\mathcal{Z}$ is divided into $K$ subsets represented by the $K$ centers. Inside each subset, the composition function (10) is smooth. The non-smoothness of (10) occurs only at the intersection of any two subsets. We denote the union of all such intersections as $S_A$. Pick any two cluster centers. Then, the points with the same distance from the two centers form a hyperplane of dimension $n + m - 1$. There exist finite $K(K - 1){\text{/2}}$ hyperplanes whose union is denoted as $S_B$. Clearly, we have that ${S_A} \subseteq {S_B}$. From the measure theory \citep{blaas2019localised}, $S_B$ has a zero measure in an ($n+m$)-dimensional space. Therefore, the set $S_A$ has a zero measure. $\Box$

Note that the Lebesgue measure of discontinuous point set is also zero as smoothness implies continuity. 

\noindent
\textbf{Lemma 3.}  The predictor $\hat f$ in (10) is bounded. Furthermore, the discontinuities of ${\hat f}$ in $S_A$ have a finite gap. 

\noindent
\textbf{Proof. }From the proof of Lemma 2, the discontinuity only occurs when a query point is divided into two or more clusters simultaneously. From (10) and (11), we have that
\begin{equation}
\hat f(w) = \sum\nolimits_{i = 1}^{N_\mathcal{D}^j} {\omega _i^j\phi _i^j(w)}  \leqslant {\sigma ^2}\sum\nolimits_{i = 1}^{N_\mathcal{D}^j} {\left| {\omega _i^j} \right|} ,\;\exists j \in I_1^K.
\end{equation}

Hence, the gap is upper bounded by the finite number
\begin{equation}
{f_d} = \mathop {\max }\limits_{1 \leqslant {j_1},{j_2} \leqslant K} \left\{ {{\sigma ^2}\left( {\sum\nolimits_{k = 1}^{N_\mathcal{D}^{{j_1}}} {\left| {\omega _k^{{j_1}}} \right|}  + \sum\nolimits_{k = 1}^{N_\mathcal{D}^{{j_2}}} {\left| {\omega _k^{{j_2}}} \right|} } \right)} \right\}.
\end{equation}
The proof of Lemma 3 is completed. $\Box$

It is noticed that Lemmas 2 and 3 provide the foundations for using the gradient-based optimization methods to solve the FHOCP of the LEMPC scheme proposed later.

To evaluate the propagation of errors over the horizon to ensure the robustness of LEMPC, we posteriorly reckon the Lipschitz constant $L_j$ in $\mathcal{Z}_j$ using the 2-norm of the Jacobian matrix $H \in {R^{n \times n}}$  of $\hat f$ on $x$, i.e., 
\begin{subequations}
\begin{equation}
{L_j} = \mathop {\max }\limits_w \left\| H \right\|
\end{equation}
\begin{equation}
{\rm{s}}{\rm{.t}}{\rm{. }}{\left\| {w - {z_j}} \right\|^2} \le {\left\| {w - {z_{{j_1}}}} \right\|^2},1 \le j,{j_1} \le K,{j_1} \ne j.
\end{equation}
\begin{equation}
w \in {\cal Z}.
\end{equation}
\end{subequations}
The optimization problem (18) is well-defined as the predictor (10) is smooth within $\mathcal{Z}_j$ by Lemma 2. Further, to ensure the robustness property of LEMPC, the upper bound $\bar \mu $ of error $\mu  = \left\| {f(w) - \hat f(w) - \delta } \right\|$ needs to be analyzed. In what follows, two results provide different upper bound estimates of $\bar \mu $. 

\noindent
\textbf{Assumption 1} The system $f$ is Lipschitz continuous with constant $L_f$, and the disturbance bound  $\bar \delta$ is known. 

Assumption 1 is a common assumption in KI-class methods \citep{calliess2014conservative, manzano2021componentwise, zheng2022nonparameteric}, used to infer the deterministic upper bound of $\mu$ as follows. 

\noindent
\textbf{Theorem 1} Consider the system (1) with Assumption 1, and the CKLR predictor. For any query point $w$ in cluster $j$, there is 
\begin{equation}
\bar \mu  \le ({L_f} + {L_j})\left\| {w - {w^*}} \right\| + 2\bar \delta  + {\bar \delta _s}
\end{equation}
where ${w^*} \in {{\cal D}_w}$ is the nearest point of $w \in {{\cal Z}_j}$.
  
\noindent
\textbf{Proof. }For any $w$, we have that
\begin{equation}
\begin{array}{l}
f(w) - \hat f(w) \le f({w^*}) + {L_f}\left\| {w - {w^*}} \right\| - \hat f({w^*}) + {L_j}\left\| {w - {w^*}} \right\|\\
 \le f({w^*}) + {L_f}\left\| {w - {w^*}} \right\| - {y^*} + {L_j}\left\| {w - {w^*}} \right\| + {{\bar \delta }_s}\\
 \le ({L_f} + {L_j})\left\| {w - {w^*}} \right\| + \bar \delta  + {{\bar \delta }_s}
\end{array}
\end{equation}
where the $y^{*}$ is the output related to $w^{*}$ in $\cal D$. Hence, 
\begin{equation}
\bar \mu  \le \mathop {{\rm{max}}}\limits_w \left\| {f(w) - \hat f(w)} \right\| + \bar \delta  \le ({L_f} + {L_j})\left\| {w - {w^*}} \right\| + 2\bar \delta  + {\bar \delta _s},
\end{equation}
\noindent
The proof of Theorem 1 is completed. $\Box$

Similar results are obtained in \citet{calliess2020lazily, calliess2018nonlinear, calliess2014conservative, manzano2019output,manzano2021componentwise,blaas2019localised,manzano2020robust,zheng2022nonparameteric}, where the prediction error converges to a constant as the data increases. However, the Lipschitz constant of a black-box model and the boundary value $\bar \delta $ are still two open issues in learning models. Here, we make probabilistic assumptions about $\mu$ and progress analysis from a test set without any assumption of system (1).

\noindent
\textbf{Assumption 2.} For a given constant $a>0$ , the distribution of error $\mu$ satisfies
\begin{equation}
\Pr (\bar \mu  - a \le \mu  \le \bar \mu ) > 0
\end{equation}
\begin{figure}  % 总体流程图
	\centering
	\includegraphics[width=.85\textwidth]{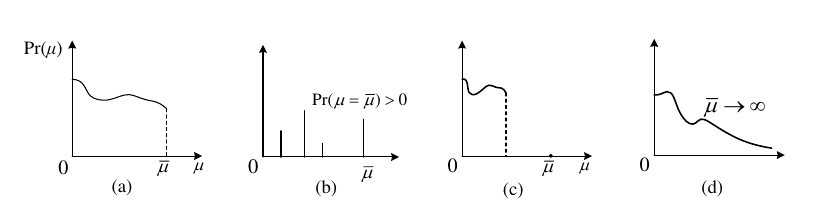}
	\caption{Visualization of four distributions. (a) bounded continuous distribution; (b) discrete distribution with nonzero probability in maximum; (c) mixed distribution with zero probability in maximum; (d) unbounded distribution. The (a) and (b) satisfy Assumption 2 whilst (c) and (d) do not.}
	\label{FIG:3}
\end{figure}
The visualization of these four distributions is shown in Fig. 3. Consider a test set ${{\cal D}_{test}}$  with $N_t$ independent and identically distributed error data with the maximum ${\bar \mu _{\max }}$. Then the probability of obtaining a conservative upper bound is provided by the following Theorem 2. 

\noindent
\textbf{Theorem 2. }Consider the system (1) with Assumption 2 and CKLR predictor. There exists a constant $p>0$  such that
\begin{equation}
\Pr (\bar \mu  \le {\bar \mu _{\max }} + a) > 1 - {(1 - p)^{{N_t}}}
\end{equation}
\textbf{Proof.} From (22), it is obtained a constant p that satisfies 
\begin{equation}
\Pr (\bar \mu  - a \le \mu  \le \bar \mu ) > p > 0
\end{equation}
Then, we have 
\begin{equation}
\begin{array}{l}
\Pr (\bar \mu  - a < {\mu _{\max }}) = 1 - \Pr ({\mu _{\max }} \le \bar \mu  - a)\\
 = 1 - \Pr {(\mu  \le \bar \mu  - a)^{{N_t}}} = 1 - {\left( {1 - \Pr (\mu  > \bar \mu  - a)} \right)^{{N_t}}}\\
 > 1 - {(1 - p)^{{N_t}}}
\end{array}
\end{equation}
The proof of Theorem 2 is completed.  $\Box$ 

\section{LEMPC ALGORITHM}
This section uses the bounds of learning errors provided by Theorems 1 and 2 to design the new LEMPC scheme. For simplicity, the bounds in Theorems 1-2 are unified as $\bar \mu $. Due to the clustering technique, the predictor is discontinuous on the clustering interfaces, leading to traditional Robust MPC algorithms \citep{defeng2019input,zhang2022robust,manzano2020robust} being unable to be used directly. For this reason, we extend the method in \citet{lazar2009predictive} to further consider mixed constraint (2) to guarantee robustness. 
\subsection{ Constraint-tightening method}
Consider the CKLR predictor ${x_{k + 1}} = \hat f({x_k},{u_k})$, where the function $\hat f$ is Lipschitz continuous about $L_j$ in each subset ${{{\cal Z}_j}}$ by Lemma 2. The maximum propagation of learning errors at prediction step $i$ is computed as 
\begin{subequations}
\begin{equation}
{\cal L}_i^{\bar \mu } = \left\{ {\eta  \in {R^n}\left| {\left\| \eta  \right\| \le \sum\nolimits_{p = 0}^{i - 1} {{L^p}\bar \mu } } \right.} \right\}.
\end{equation}
\begin{equation}
L = \max {L_j},j \in I_1^K
\end{equation}
\end{subequations}
Correspondingly, the nominal constraints are designed as 
\begin{equation}
{\bar {\cal Z}_i} = \bigcup\limits_{j = 1}^K {\left( {{{\cal Z}_j} \ominus {\cal L}_i^{\bar \mu } \times \mathbf{0}} \right)} ,\forall i \in I_0^N
\end{equation}
where $\mathbf{0} \in {R^m}$. Then, the LEMPC algorithm is designed as follows. 

\subsection{LEMPC algorithm design}
Consider the economically optimal steady-state point of the predictor of (1), computed by 
\begin{equation}
({x_s},{u_s}) = \mathop {\arg \min }\limits_{(x,u) \in {\cal Z}} \;{L_e}(x,u),{\rm{  s}}{\rm{.t}}{\rm{.}}x = \hat f(x,u).
\end{equation}

\noindent
Note that $({x_s},{u_s})$ is not necessarily the optimal economic steady state of the system (1) as $\hat f \ne f$ in general. Since the measure of $S_A$ is 0, it is practically supposed that there exists a $j_0$ such that $({x_s},{u_s}) \in {\mathop{\rm int}} ({\bar {\cal Z}_{{j_0}}})$. Without loss of generality, it is assumed that $({x_s},{u_s})$ is unique and equals the origin via coordinate transformation. 

Consider the finite horizon economic cost function (5) with the prediction horizon $N>0$. In general, optimizing direct economic function may lead to instability \citep{diehl2010lyapunov,rawlings2017model}. Hence, an auxiliary positive definite function is introduced, i.e.,
\begin{equation}
{J_a}\left( {{x_k},{\mathbf{u}_k}} \right) = {E_a}({x_{N|k}}) + \sum\nolimits_{i = 0}^{N - 1} {{L_a}\left( {{x_{i|k}},{u_{i|k}}} \right)} ,
\end{equation}
where the auxiliary functions $L_a$ and $E_a$ are positive definite concerning $({x_s},{u_s})$ and $x_s$, respectively. And $L_a$ satisfies that
\begin{equation}
\left\| {{L_a}\left( {{x_1},u} \right) - {L_a}\left( {{x_2},u} \right)} \right\| \le {c_L}\left\| {{x_1} - {x_2}} \right\|,\forall ({x_1},u),({x_2},u) \in {\cal Z}
\end{equation}
with $c_L>0$. Furthermore, there exist some ${{\cal K}_\infty }$ functions $\gamma_0$, $\gamma_1$, and $\gamma_2$ such that for all $(x,u) \in {\cal Z}$,
\begin{equation}
{\gamma _0}\left( {\left\| x \right\|} \right) \le {L_a}\left( {x,u} \right) \le {\gamma _1}\left( {\left\| x \right\|} \right) + {\gamma _2}\left( {\left\| u \right\|} \right)
\end{equation}
The $L_a$ and $E_a$ can be chosen as quadratic positive definite functions in stabilizing MPC schemes \citep{defeng2019input,manzano2019output}.

For the state $x_k$ at time $k \ge 0$, the FHOCP is formulated as
\begin{subequations}
\begin{equation}
V_e^*\left( {{x_k}} \right) = \mathop {\min }\limits_{{\mathbf{u}_k}} {J_e}\left( {{x_k},{u_k}} \right),
\end{equation}
\begin{equation}
\;{\rm{s}}{\rm{.t}}{\rm{.}}\;\;{x_{i + 1|k}} = \hat f\left( {{x_{i|k}},{u_{i|k}}} \right),
\end{equation}
\begin{equation}
{x_{0|k}} = {x_k},
\end{equation}
\begin{equation}
{\rm{(}}{u_{i|k}},{x_{i|k}}{\rm{)}} \in {\bar {\cal Z}_i},i \in I_{\rm{0}}^{N - 1}{\rm{.}}
\end{equation}
\begin{equation}
{x_{N|k}} \in {X_f},
\end{equation}
\begin{equation}
{J_a}\left( {{x_k},{\mathbf{u}_k}} \right) \le \Pi ({x_k}).
\end{equation}
\end{subequations}
where $V_e^*({x_k})$ is the optimal value function of economic cost (5) with the optimal solution $ \mathbf{u}_k^* = \{ u_{0|k}^*,u_{1|k}^*...\;u_{N - 1|k}^*\} $, and the constraints (32b)-(32f) are the predictor, initial condition, nominal constraints, terminal state constraint, and contraction constraint, respectively. From the receding horizon control principle, the first element of $\mathbf{u}_k^*$ is used to feedback control system, and the closed-loop dynamic is 
\begin{equation}
{x_{k + 1}} = f({x_k},\,u_{0|k}^*) + {\delta _k},\quad k \in {I_{ \ge 0}}
\end{equation}
Consider the predictor ${x_{k + 1}} = \hat f({x_k},{u_k})$ and define the sets  
\begin{equation}
{X_u} = \{ x \in {R^n}:\left( {x,{K_{\hat f}}(x)} \right) \in {\bar {\cal Z}_{N - 1}}\} 
\end{equation}
\begin{equation}
{X_p} = \{ x \in {R^n}:{E_a}(x) \le {\alpha _p}\}  \subseteq {X_u},
\end{equation}
with the factor ${\alpha _p} \ge 0$. The following classic terminal assumption is used to guarantee feasibility and ISS.

\noindent
\textbf{Assumption 3.} There exists a stabilizing local feedback law $u = {K_{\hat f}}(x)$ defined in $X_p$ such that ${K_{\hat f}}(0) = 0,\left( {x,{K_{\hat f}}(x)} \right) \in {{\cal Z}_{{j_0}}}$, and
\begin{equation}
{E_a}\left( {\hat f\left( {x,{K_{\hat f}}(x)} \right)} \right) - {E_a}(x) \le  - {L_a}\left( {x,{K_{\hat f}}(x)} \right),\forall x \in {X_p}.
\end{equation}
Moreover, there exists a ${{\cal K}_\infty }$-function $K_u$ such that the optimal control sequence $\mathbf{u}_k^*$ satisfies that
\begin{equation}
\left\| {u_{i|k}^*} \right\| \le {K_u}\left( {\left\| x \right\|} \right),{\rm{ }}i \in I_1^{N - 1},\forall x \in {X_p}.
\end{equation}

\noindent
\textbf{Remark 3.} The inequality (36) implies that ${E_a}(x)$  constructs a local Lyapunov function of system ${x_{k + 1}} = \hat f{\rm{(}}x,{K_{\hat f}}(x){\rm{)}}$ in $X_p$, whilst (37) indirectly ensures that $V_e^*({x_k})$ has a ${\cal K}$-class function upper bound (compared with Assumption 2.4 in \citet{faulwasser2018economic}). Note that the difficult verification of (37) can be bypassed by switching to the stabilized control law ${K_{\hat f}}(x)$ in $X_p$. 

Now, we consider the terminal set $X_f$ of the predictor. Define ${X_{\hat f}} = \{ x \in {R^n}:{E_a}(x) \le {\alpha _N}\} $ such that 
\begin{equation}
\hat f\left( {x,{K_{\hat f}}(x)} \right) \in {X_f},\;\forall x \in {X_p}.
\end{equation}
where the number $0 < {\alpha _N} \le {\alpha _p}$. Note that $\alpha_N$ always exists as $X_p$ is nonempty.

To design the contraction constraint (32f), we consider the following FHOCP
\begin{subequations}
\begin{equation}
V_a^*({x_k}) = \mathop {\min }\limits_{{\mathbf{u}_k}} {J_a}\left( {{x_k},{\mathbf{u}_k}} \right),
\end{equation}
\begin{equation}
s.t.{\rm{ (}}32b{\rm{) - }}(32e),
\end{equation}
\end{subequations}
where the optimal value function $V_a^*({x_k}) = {J_a}({x_k},\mathbf{u}_k^a)$ with the optimal solution $\mathbf{u}_k^a = \{ u_{0|k}^a,u_{1|k}^a...\;u_{N - 1|k}^a\}$. Then the value $\Pi ({x_k})$ in (32f) is calculated as \begin{equation}
\Pi ({x_k}) = (1 - \alpha )\left( {\bar e\chi  + V_a^e({x_{k - 1}})} \right) + \alpha V_a^*({x_k}),
\end{equation}
where the auxiliary value function
\begin{equation}
V_a^e({x_{k - 1}}) = {J_a}({x_{k - 1}},\mathbf{u}_{k - 1}^*)
\end{equation}
and the factors $0 < \alpha  \le 1$, and 
\begin{equation}
\chi  = {c_E}{L^{N - 1}} + {c_L}{{({L^{N - 1}} - 1)} \mathord{\left/
 {\vphantom {{({L^{N - 1}} - 1)} {(L - 1)}}} \right.
 \kern-\nulldelimiterspace} {(L - 1)}}.
\end{equation}
where $c_E>0$ is the local Lipschitz constant of $E_a$, i.e. $||{E_a}({x_1}) - {E_a}({x_2})|| \le {c_E}||{x_1} - {x_2}||,\forall {x_1},{x_2} \in {X_p}$. From \citet{he2016economic, defeng2019input}, the $\alpha$ can balance the convergence speed and economic performance of the closed-loop system.

Define two constants $L_a^{\max },E_a^{\max }$ as
\begin{equation}
L_a^{\max } = \mathop {\;\max \;}\limits_{(x,u) \in {\cal Z}} {L_a}\left( {x,u} \right),
\end{equation}
\begin{equation}
E_a^{\max } = \mathop {\;\max \;}\limits_{(x,u) \in {\cal Z}} {E_a}\left( {x,u} \right).
\end{equation}
Then, the initialization of $\Pi ({x_k})$ can be set as 
\begin{equation}
\Pi ({x_0}) = NL_a^{\max } + E_a^{\max } + 1
\end{equation}
such that (32f) is inactive at $k=0$. 

The whole LEMPC algorithm is summarized as follows. 

\begin{center}
\begin{tabularx}{\linewidth}{X}
\toprule
\textbf{Algorithm 2: LEMPC Algorithm} \\
\midrule
\textbf{Offline: } \\
Step 1. Run CKLR to derive $L_j$ for each ${\cal Z}_j$ and then compute ${\bar {\cal Z}_j}$.   \\
Step 2. Compute $(x_s, u_s)$ and design $L_a(x,u)$ and $E_a(x)$.  \\
\textbf{Offline: } \\
Step 3. (initialization) Measure $x_k$ at initial time $k=0$ and initialize $\Pi ({x_0})$ as in (45); Solve the problem (32) and implement the first element of $\mathbf{u}_0^{*}$ to the system (1), let $k=k+1$.   \\
Step 4. Solve the problem (39) to update $\Pi ({x_k})$.  \\
Step 5. Solve the problem (32) to derive $\mathbf{u}_k^{*}$ and implement its first element to the system (1).\\ 
Step 6. Let $k=k+1$, and return to Step 4.\\
\bottomrule
\end{tabularx}
\end{center}

\noindent
\textbf{Remark 4.} Note that the feasible control solutions to (39) can be used to compute $\Pi ({x_k})$ in the price of convergence speed of state trajectories \citep{defeng2019input}. Here, we use the optimal value function $V_a^*({x_k})$  to derive $\Pi ({x_k})$, which accelerates the convergence speed but degrades the economic performance. To resolve this conflict, one can decrease $\alpha$ to improve the economic performance of the system to some extent. In addition, our LEMPC could recover the stabilization effect of stabilized MPC under $\alpha=1$.

\subsection{Stability analysis}
At initial time $k=0$, we define the $N$-step feasible domain as
\begin{equation}
\mathbf{X}_N = \left\{ x \in \mathbb{R}^n \,\middle|\, 
\begin{aligned}
&\exists \mathbf{u} = \{ u_0, u_1, \dots, u_{N-1} \}, \\
&x_{i+1} = \hat{f}(x_i, u_i), \; x_0 = x, \\
&(x_i, u_i) \in \bar{\mathcal{Z}}_i, \; i \in I_0^{N-1}, \\
&x_N \in X_f
\end{aligned}
\right\}.
\end{equation}
The main theoretical results are presented as follows.

\noindent
\textbf{Theorem 3.} Under Assumptions 1 and 3, the problems (32) and (39) are recursively feasible in the robust invariant set $\mathbf{X}_N$ if the error bound $\bar \mu $ satisfies that
\begin{equation}
{c_E}{L^{N - 1}}\bar \mu  + {\alpha _N} \leqslant {\alpha _p}.
\end{equation}

\noindent
\textbf{Proof.} Let $\mathbf{u}_{k - 1}^*$ be the optimal solution to (32) with $\forall {x_k}_{ - 1} \in {{\mathbf{X}}_N}$ at time $k-1$. We first find feasible solutions to (39) at $k$ and then establish the theorem. 
To this end, a candidate solution to (39) at $k$ is selected as 
\begin{equation}
{\mathbf{\tilde u}_k} = \{ u_{1|k - 1}^*...\;u_{N - 1|k - 1}^*,u_{N|k - 1}^*\} ,
\end{equation}
where $u_{N|k - 1}^* = {K_{\hat f}}(x_{N|k - 1}^*)$ and the corresponding state trajectories are denoted as ${x_i}_{|k}$ for $i \in I_1^{N - 1}$ . From Lemma 11 in \citet{lazar2009predictive}, we have that $\left\| {{x_{i|k}} - x_{i + 1|k - 1}^*} \right\| \le {L^i}\bar \mu$ since $\left\| {{x_k} - x_{1|k - 1}^*} \right\| \le \bar \mu $. Moreover, since $(x_{i + 1|k - 1}^*,u_{i + 1|k - 1}^*) \in \bar {\cal Z}_{i + 1}^{},i \in I_1^{N - 1}$, we have that
\begin{equation}
({x_{i|k}},u_{i + 1|k - 1}^*) \in {\bar {\cal Z}_{i + 1}} \oplus \left( {\left\{ {\eta  \in {R^n}\left| {\left\| \eta  \right\| \le {L^i}\bar \mu } \right.} \right\} \times 0} \right) \subset {\bar {\cal Z}_i},
\end{equation}
\noindent
which implies the (32d) at $k$. Combining the positive-definiteness and Lipschitz continuity of ${E_a}(x)$, it is obtained that 
\begin{equation}
{E_a}({x_{N - 1|k}}) \le {E_a}(x_{N|k - 1}^*) + {c_E}{L^{N - 1}}\bar \mu  \le {\alpha _N} + {c_E}{L^{N - 1}}\bar \mu  \le {\alpha _p}.
\end{equation}
\noindent
Then, from Assumption 3, it is obtained that ${x_{N|k}} \in {X_f}$. Namely, the constraint (32e) is satisfied at $k$. Hence, the sequence (48) is a feasible solution to (39). 

Now, we prove that (32) has at least one feasible solution at $k$. Notice that since (39) is feasible at $k$, it has an optimal solution $\mathbf{u}_k^a =\{ {u_{0|k}^a,u_{1|k}^a...\;u_{N - 1|k}^a}\}$ at the time $k$. From the optimality principle, we have that
\begin{equation}
V_a^*({x_k}) \le {J_a}\left( {{x_k},{\mathbf{{\tilde u}}_k}} \right).
\end{equation}
\noindent
which yields 
\begin{equation}
\begin{array}{l}
V_a^*({x_k}) - V_a^e({x_{k - 1}}) \le {J_a}( {{x_k},{\mathbf{{\tilde u}}_k}} ) - V_a^e({x_{k - 1}})\\
 = \sum\limits_{i = 0}^{N - 1} {\left( {{L_a}{\rm{(}}{x_{i|k}},u_{i + 1|k - 1}^*{\rm{)}} - {L_a}{\rm{(}}{x_{i|k - 1}},u_{i|k - 1}^*{\rm{)}}} \right)}  + {E_a}({x_{N|k}}) - {E_a}({x_{N|k - 1}})\\
 = \sum\limits_{i = 0}^{N - 2} {\left( {{L_a}({x_{i|k}},u_{i + 1|k - 1}^*) - {L_a}({x_{i + 1|k - 1}},u_{i + 1|k - 1}^*)} \right)}  \\
 + \left( {{E_a}({x_{N - 1|k}}) - {E_a}({x_{N|k - 1}}) + {E_a}({x_{N|k}}) - {E_a}({x_{N - 1|k}})} \right) + {L_a}({x_{N - 1|k}},u_{N|k - 1}^*) - {L_a}({x_{0|k - 1}},u_{0|k - 1}^*)
\end{array}
\end{equation}
From (30), it is derived that 
\begin{equation}
{L_a}( {{x_{i|k}},u_{i + 1|k - 1}^*} ) - {L_a}( {{x_{i + 1|k - 1}},u_{i + 1|k - 1}^*} ) \le {c_L}{L^i}\bar \mu ,
\end{equation}
\begin{equation}
{E_a}({x_{N - 1|k}}) - {E_a}({x_{N|k - 1}}) \le {c_E}{L^{N - 1}}\bar \mu .
\end{equation}
\noindent
Substituting (36), (42) and (53)-(54) into (52), we have that
\begin{equation}
V_a^*({x_k}) - V_a^e({x_{k - 1}}) \le  - {L_a}( {{x_{0|k - 1}},u_{0|k - 1}^*} ) + \bar \mu \chi .
\end{equation}
\noindent
which implies that 
\begin{equation}
V_a^*({x_k}) \le \left( {1 - \alpha } \right)\left( {\bar \mu \chi  + V_a^e({x_{k - 1}})} \right) + \alpha V_a^*({x_k}) = \Pi ({x_k}).
\end{equation}

That is, the constraint (32f) is satisfied at time $k$. Hence, the optimal solution $\mathbf{u}_k^a$ to (39) is a feasible solution to (32) at $k$ and then ${x_k} \in {{\bf{X}}_N}$ for $\forall {x_k}_{-1} \in {{\bf{X}}_N}$. This implies that the problem (32) has recursive feasibility in the robust invariant set ${{\bf{X}}_N}$. 

By repeating the proof, we obtain the optimal solution to (32) at time $k$, which guarantees at least one feasible solution to (39) at $k+1$. As a consequence, the recursive feasibility of the LEMPC algorithm, i.e., (32) and (39), is ensured in the robust invariant set ${{\bf{X}}_N}$. $\Box$

\noindent
\textbf{Theorem 4.} Under Assumptions 1 and 3, the closed-loop system (33) admits ISS in the robust invariant set ${{\bf{X}}_N}$ if the problem (32) is feasible at $k=0$. 

\noindent
\textbf{Proof.} Since the problem (32) is feasible at $k=0$, its optimal solution always exists at each time according to Theorem 3. Let $\mathbf{u}_k^*$ be the optimal solution to (32) at time $k$ and select the value function $V_a^e(x)$ in (41) as the candidate of ISS Lyapunov functions of the closed-loop system in (33). 

First, consider the bounds of function $V_a^e(x)$. It is true that
\begin{equation}
V_a^e(x) \ge {L_a}{\rm{(}}x,u) \ge {\gamma _0}(||x||),\;\forall (x,u) \in {\cal Z}
\end{equation}
Consider the equilibrium point $({x_s},{u_s}) \in {\mathop{\rm int}} ({\bar {\cal Z}_{{j_0}}})$. From Lemma 2, there exists a neighborhood of $({x_s},{u_s})$ such that the predictor is smooth in the neighborhood. Therefore, there exists a ${\cal K}$-function $\vartheta $ such that 
\begin{equation}
\left\| {\hat f{\rm{(}}x,u{\rm{)}} - \hat f{\rm{(}}z,v{\rm{)}}} \right\| \le \vartheta \left( {\left\| {{\rm{(}}x,u{\rm{)}} - {\rm{(}}z,v{\rm{)}}} \right\|} \right),
\end{equation}
where pairs $(x,u)$ and $(z,v)$ are in the neighborhood. From the definition of $V_a^e({x_k})$ and combining (31), (58), we have that
\begin{equation}
\begin{array}{l}
V_a^e({x_k}) \le {c_E}\left\| {x_{N|k}^*} \right\| + \sum\nolimits_{i = 0}^{N - 1} {{\gamma _1}(\left\| {x_{i|k}^*} \right\|) + {\gamma _2}(\left\| {u_{i|k}^*} \right\|)} \\
 \le {c_E}\left\| {\hat f\left( {x_{N - 1|k}^*,u_{N - 1|k}^*} \right)} \right\| + {\gamma _1}\left( {\left\| {x_{0|k}^*} \right\|} \right) + \sum\limits_{i = 0}^{N - 1} {{\gamma _2}\left( {\left\| {u_{i|k}^*} \right\|} \right)}  + \sum\limits_{i = 0}^{N - 1} {{\gamma _1}\left( {\left\| {\hat f\left( {x_{i|k}^*,u_{i|k}^*} \right)} \right\|} \right)} \\
 \le {c_E}\left( {{\varpi _N}\left( {{x_k}} \right)} \right) + N{\gamma _2}\left( {{K_u}\left( {\left\| {{x_k}} \right\|} \right)\;} \right) + \sum\limits_{i = 0}^{N - 1} {{\gamma _1}\left( {{\varpi _i}\left( {\left\| {{x_k}} \right\|} \right)} \right){\rm{ }}} \\
 = {{\bar \alpha }_1}({x_k})
\end{array}
\end{equation}
where ${\varpi _i},\forall i \in I_0^N$ and ${\bar \alpha _2}$ are ${{\cal K}_\infty }$-functions. The functions ${\varpi _0}$ is identify mapping and ${\varpi _i} = (1 + {K_u})\vartheta ({\varpi _{i - 1}}),\forall i \le I_1^N$. On the other hand, we have 
\begin{equation}
V_a^e(x) \le NL_a^{\max } + E_a^{\max },\forall x \in {X_N}
\end{equation}
Combining (59) and (60), there exists a ${{\cal K}_\infty }$-function ${\alpha _2}$ such that $V_a^e(x) \le {\alpha _2}\left( {\left\| x \right\|} \right),\forall x \in {X_N}$ by Proposition B.25 in \citet{rawlings2017model}. 

Now we consider the difference of $V_a^e(x)$ along the trajectories of the closed-loop system (33). We have
\begin{equation}
\begin{array}{l}
V_a^e({x_k}) - V_a^e({x_{k - 1}}) \le \Pi ({x_k}) - V_a^e({x_{k - 1}})\\
 = (1 - \alpha )\bar \mu \chi  + \alpha \left( {V_a^*({x_k}) - V_a^e({x_{k - 1}})} \right)\;\\
 \le (1 - \alpha )\bar \mu \chi  + \alpha \left( { - {L_a}( {{x_{0|k - 1}},u_{0|k - 1}^*} ) + \bar \mu \chi } \right)\\
 \le  - \alpha {\gamma _0}\left( {\left\| x \right\|} \right) + \bar \mu \chi 
\end{array}
\end{equation}
Hence, $V_a^e(x)$ is an ISS-Lyapunov function of (33). Finally, Theorem 4 holds from Lemma 1. $\Box$

\noindent
\textbf{Theorem 5.} Consider the CKLR predictor of the unknown system (1). Suppose Assumptions 2 and 3 hold,  and the problem (32) is feasible at $k=0$, then there exists a constant $p>0$ such that, with the probability $1 - {(1 - p)^{{N_t}}}$, the LEMPC problems in Algorithm 2 are recursively feasible and the closed-loop system (33) admits ISS in the robust invariant set  ${{\bf{X}}_N}$. 

\noindent
\textbf{Proof.} The proof follows from Theorems 3 and 4, omitted here to save space. $\Box$ 

\section{Examples}
In this section, two examples verify the performance of the presented schemes. A nonlinear function is used to evaluate the CKLR learning method and a CSTR is adopted to verify the LEMPC scheme. All optimizations are performed on a laptop with an Intel $10^{th}$ i5 processor and MatLab 2023b. The offline QCQPs are solved by Gurobi, and the FHOCPs are solved by the sequential quadratic programming algorithm. 

\subsection{A numerical example }
Consider the following unknown nonlinear mapping 
\begin{equation}
y = 0.8{(x - 10)^2} + 8\cos (u) + \delta 
\end{equation}
with constraint $(x,u) \in [0,20] \times [0,20]$ and disturbance $\delta  \in [ - 1,{\rm{ }}1]$, where 200 data points under uniform distribution are generated to learn the mapping. The clustering number is selected as $K=2$. The Lipschitz constant is estimated by (9) as $\bar L = 15.3832$  and the actual (unknown) Lipschitz constant is 17.88. The parameters in (12) are ${{\bar \delta }_s} = 1.5,{\rm{ }}{\sigma ^2} = 1,{\rm{ }}l = 5,{\rm{ }}S = 2000$ and the maximal optimization time of solving QCQP is about 37.07 s.

To verify the merits of the designed CKLR scheme, we compare our algorithm with KI in \citet{calliess2014conservative}, PKI and SPKI in \citet{manzano2019output}, SLR in \citet{maddalena2020learning}, and the standard Radial Basis Function Neural Network (RBFNN) in \citet{murphy2022probabilistic}. Note that KI is identical to PKI if $K=2$, as PKI also considers the neighbour subset. In KI, all data are directly used to infer the output, while PKI uses rectangular partitions of data, and each output of SPKI is a convex weighted combination of adjacent outputs of PKI. Here, the distance is 0.2 and the weight of each adjacent point is 0.025. The SLR directly solves the optimization problem on the whole data set with the same parameters as CKLR, whose computational time is 63.43s. RBFNN contains linear layers and automatically adjusts the neuron number to minimize mean squared error. 

\begin{figure}  
	\centering
	\includegraphics[width=.96\textwidth]{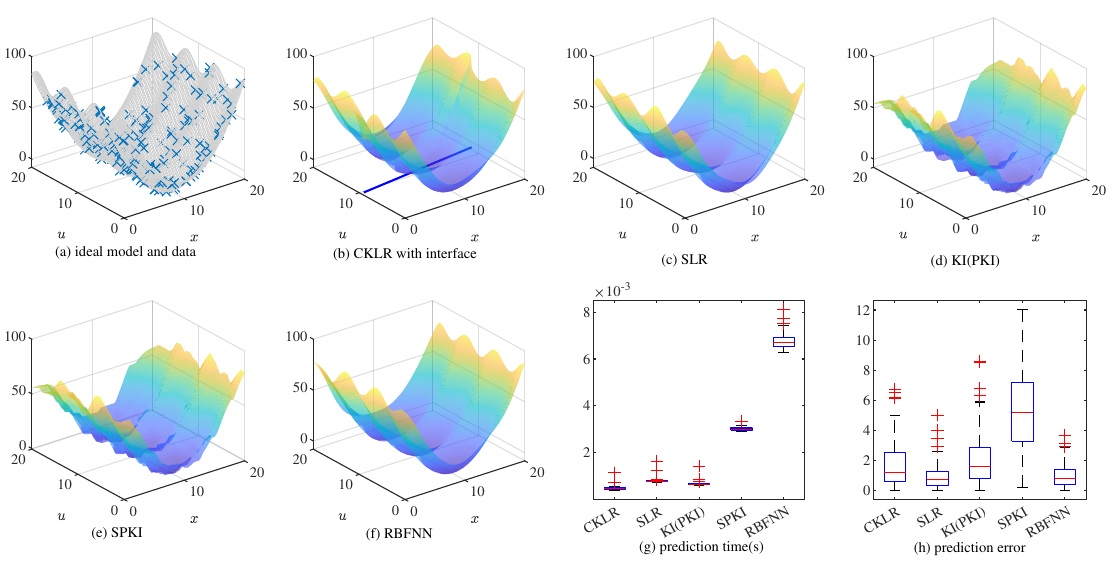}
	\caption{(a) The actual model represented by the gray grid and 200 random data points represented by blue crosses; (b) The CKLR predictor ($K=2$) and interface of clusters represented by a blue line; (c) SLR predictor; (d) KI (PKI) predictor; (e) SPKI predictor; (f) RBFNN predictor; (g) Boxplot of prediction time on 100 test points of five algorithms; (h) Boxplot of absolute prediction error on 100 test points of five algorithms.}
	\label{FIG:4}
\end{figure}

Fig.4 shows the real mapping, data, predictor profiles under five methods, accuracy, and online prediction time on 100 test points. As can be seen from Fig.4 (a)-(c), the learning results of CKLR and SLR are similar, while our CKLR has slight discontinuities at the interface of the clusters, displayed by the blue line. In Fig.4 (d)-(e), KI (PKI), SPKI all learn a typical non-smooth model. In Fig.4 (f), RBFNN fits the continuous dynamics of the system accurately. In Fig.4 (g), CKLR has the faster prediction speed, being approximately half of SLR since the data set is divided into two clusters. KI (PKI) is slightly slower as it needs to use the whole data set, and SPKI is slower than KI due to the predictions of neighboring points and convex combination calculation. In RBFNN, the linear layers and the enormous number of nodes for minimizing mean squared error significantly increase online prediction calculation. In Fig.4 (h), the kernel function makes CKLR, SLR and RBFNN have better accuracy than KI (PKI) and SPKI. SLR is more accurate as the SLR predictor uses the whole data set rather than a single clustering. Hence, more kernel functions are used in SLR. Finally, RBFNN has the best accuracy and most complex predictor among the five methods since its unlimited kernel number directly optimizes the mean squared error. Therefore, our CKLR method achieves satisfactory results regarding model smoothness, learning accuracy, and fast online prediction.

To further illustrate the merits of our CKLR in promoting higher accuracy and obtaining a smaller Lipschitz constant, we test the $R$-square and calculate the Lipschitz constant $L$ of five methods. The results are reported in Table 1. Note that the values of $L$ of our CKLR and SLR are less than RBFNN's since (12d) limits the Lipschitz constant at $S=2000$ points. The features of the five methods are summarized in Table 2. To sum up, our CKLR learning algorithm is the only one that achieves high accuracy, big data tractability, learning accuracy, and a smaller Lipschitz constant.

\begin{table}[h]
\caption{\textit{R-square} and Lipschitz constant \textit{L} of five methods}\label{tbl1}
\centering
\begin{tabular}{|l|l|l|l|l|l|}
\hline
 & CKLR & SLR & KI & SPKI & RBFNN  \\  
\hline
$R$-square & 0.9880 & 0.9899 & 0.9650 & 0.9070 & 0.9950 \\
L &  15.2809 & 15.3213 & 15.1647 & 13.6450 & 18.0589 \\
\hline
\end{tabular}
\end{table}

\begin{table}[h]
\caption{Features of five methods}\label{tbl2}
\centering
\begin{tabular}{|l|l|l|l|l|l|}
\hline
 & CKLR & SLR & KI & SPKI & RBFNN  \\  
\hline
high accuracy &  \checkmark  & \checkmark &  \ding{53}  &   \ding{53}  &  \checkmark \\
big data scalability &  \checkmark  & \ding{53}   &  \checkmark   &  \checkmark &  \checkmark   \\
fast online prediction & \checkmark & \ding{53} & \ding{53} & \checkmark  & \ding{53} \\
smaller Lipschitz constant &  \checkmark  & \checkmark  & \checkmark  & \checkmark & \ding{53} \\
\hline
\end{tabular}
\end{table}

\subsection{A CSTR example }
In this subsection, we consider the CSTR system [16, 38]
\begin{equation}
\begin{array}{l}
\frac{{d{C_A}(t)}}{{dt}} = \frac{{{q_0}}}{V}\left( {{C_{Af}} - {C_A}(t)} \right) - {k_0}{e^{ - {\rm{ }}\frac{E}{{RT(t)}}}}{C_A}(t) + {\delta _1}(t)\\
\frac{{dT(t)}}{{dt}} = \frac{{{q_0}}}{V}\left( {{T_f} - T(t)} \right) - \frac{{\Delta {H_r}{k_0}}}{{\rho {C_p}}}{e^{ - {\rm{ }}\frac{E}{{RT(t)}}}}{C_A}(t) + \frac{{UA}}{{V\rho {C_p}}}\left( {{T_c}(t) - T(t)} \right) + {\delta _2}(t)\\
\frac{{d{T_c}(t)}}{{dt}} = \frac{{{T_c}(t) - {T_r}(t)}}{\tau } + {\delta _3}(t)
\end{array} 
\end{equation}
\noindent
where the state $x = \left[ {{C_A},T,{T_c}} \right]$ and input $u=T_r$. The $ {\delta _i} \in [ - 2 \times {10^{ - 5}},{\rm{ }}2 \times {10^{ - 5}}]$ is the additive disturbance for $i \in I_1^3$. The constraints are $0.2 \le {C_A} \le 0.8,{\rm{ }}300 \le T \le 370,{\rm{ }}280 \le {T_C} \le 350,{\rm{ }}280 \le {T_r} \le {\rm{ }}360$ . Other model parameters are given in Table 3. Select the sampling time $T_s=0.5$ min for Euler discretization and consider the economic criterion 
\begin{equation}
{L_e}(x,u) = u + {\left( {{x_2} - 345} \right)^2}.
\end{equation}
The control goal of the CSTR is to minimize energy consumption while achieving stability of some economically steady-state points in the presence of constraints.  

\begin{table}[h]
\caption{Model parameters of the CSTR}\label{tbl3}
\centering
\begin{tabular}{|l|l|l|l|}
\hline
Parameter & Definition & Value & Units  \\  
\hline
$q_0$ &  input flow  & 10 &  l/min  \\
$V$ &  liquid volume & 120   &  l    \\
$k_0$ & frequency constant & $6 \times {10^{10}}$ & l/min  \\
$E/R$ & Arrhenius constant & 9750 & K  \\
$- \Delta {H_r}$ &  enthalpy & $1 \times {10^{4}}$ & J/mol  \\
$U \cdot A$ &  transfer coefficient & $5 \times {10^{4}}$ & (J·K)/min    \\
$\rho$ & density & 1100 & g/l  \\
$C_p$ & specific heat & 0.25  & (J·K)/g  \\
$\tau$ & time constant  & 1.5 &  min  \\
$V_{Af}$ &  $C_A$ in input flow  & 1  & mol/l    \\
$T_f$ & temperature & 345 & K \\
\hline
\end{tabular}
\end{table}

In the offline CKLR learning stage, we normalize the model and set $K=81$, $\sigma_2=1$, $l=1$ and $S=300$. The upper bound of componentwise Lipschitz constants of (63) is calculated as ${\rm{[0}}{\rm{.9994, 0}}{\rm{.7108, 0}}{\rm{.7675]}}{\rm{.}}$ The maximum time Gurobi takes to solve (12) is 0.6317 s. Selecting PKI and SPKI \citep{manzano2019output} as benchmarks, the online computation time and prediction accuracy of CKLR are evaluated in extra 300 test points. Here, PKI evenly divides each dimension into three parts to construct 81 hyperrectangles for partitioning data. The results are shown in Fig. 5. Even though the data meets the uniform distribution, which is suitable to PKI and SPKI, our CKLR still has better accuracy than both PKI and SPKI, and the maximal test error of CKLR predictor is 0.0077. The prediction time of CKLR is slightly increased compared with that of PKI and SPKI due to the explicit calculation of the nonlinear smooth kernel function in (11).

For calculating the Lipschitz constant $L$ of the CKLR predictor, we do not optimize (18) directly as it is a complex nonconvex problem. Thus, we randomly generate $1 \times {10^4}$ data in constraint ${\cal Z}$ to explicitly reckon the 2-norm of the Jacobian matrix and take the maximum as estimation $L=1.1376$. 

In the control stage, the economically optimal steady-state point of the predictor is calculated as ${x_s} = (0.8588,{\rm{ }}0.6354,{\rm{ }}0.9202)$ and $u_s=0.8052$, with a corresponding economic cost of 64.6888. The Jacobian matrixes of the CKLR predictor at $x_s$ are computed as 
\begin{equation}
A = \left[ {\begin{array}{*{20}{c}}
{0.9445}&{ - 0.0385}&{0.0013}\\
{ - 0.0013}&{0.3501}&{0.6178}\\
{7.9 \times {{10}^{ - 4}}}&{ - 0.0075}&{0.6562}
\end{array}} \right],B = \left[ \begin{array}{l}
0.0057\\
0.0014\\
0.3788
\end{array} \right]{\rm{.}}
\end{equation}
Moreover, select the auxiliary cost function as
\begin{equation}
{J_a}({x_k},{u_k}) = \sum\limits_{i = 0}^{N - 1} {\left( {\bar x_{i|k}^{\rm{T}}Q\bar x_{i|k}^{} + \bar u_{i|k}^{\rm{T}}R{{\bar u}_{i|k}}} \right)}  + \bar x_{N|k}^{\rm{T}}P{\bar x_{N|k}},
\end{equation}
where $\bar x = x - {x_s},\bar u = u - {u_s}$, $R = 1,Q = 0.5I$ with unit matrix $I$. The matrix $P = \left[ {\begin{array}{*{20}{c}}
{4.5871}&{ - 0.2445}&{ - 0.2956}\\
{ - 0.2445}&{0.5836}&{0.1630}\\
{ - 0.2956}&{0.1630}&{1.3066}
\end{array}} \right]$ is obtained by solving the discrete-time Riccati equation.

Let $N=6$ be the prediction horizon. The terminal parameters are calculated as $\alpha_p=0.0194$ and $\alpha_N=0.0071$, and the error upper bound is 0.00881 using (47). From Fig.5, only the CKLR algorithm can achieve fast prediction and satisfy the accuracy requirement. 

To evaluate the control effect of the proposed LEMPC, four different initial states [0.8, 0.57, 0.6], [0.85, 0.3, 0.7], [0.83, 0.6, 0.4], [0.81, 0.8, 0.1] are picked and let $\alpha =0.97$. The state trajectories of system (63) in closed-loop with the LEMPC under the CKLR predictor (solid and lines) and ideal predictor (dash-dotted lines) are shown in Fig.6. Note that the ideal predictor is unavailable in practice. From Fig. 6, one can see that all state trajectories driven by both predictors converge to the terminal set $X_f$ while satisfying the constraints. This implies that the CKLR method reconstructs an accurate model, and the proposed LEMPC can perfectly achieve the robust control performance of uncertain systems.

\begin{figure}  
	\centering
	\includegraphics[width=.5\textwidth]{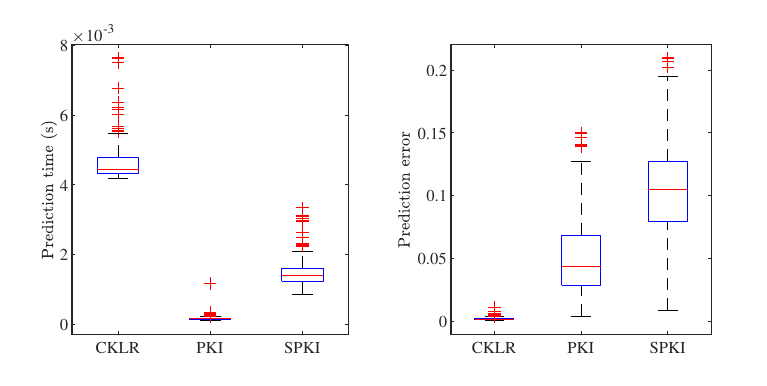}
	\caption{Comparison of prediction time and error }
	\label{FIG:5}
\end{figure}

\begin{figure}  
	\centering
	\includegraphics[width=.5\textwidth]{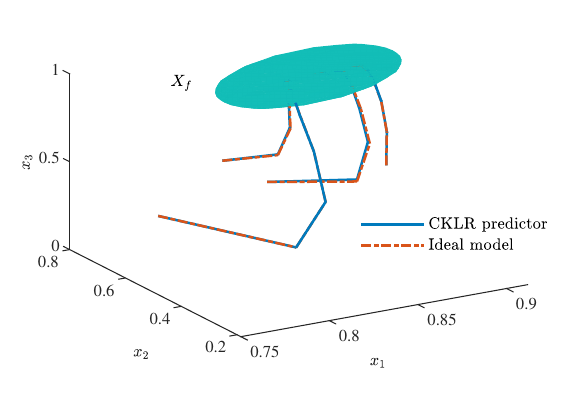}
	\caption{State trajectories of different initial points}
	\label{FIG:6}
\end{figure}

\begin{figure}  
	\centering
	\includegraphics[width=.5\textwidth]{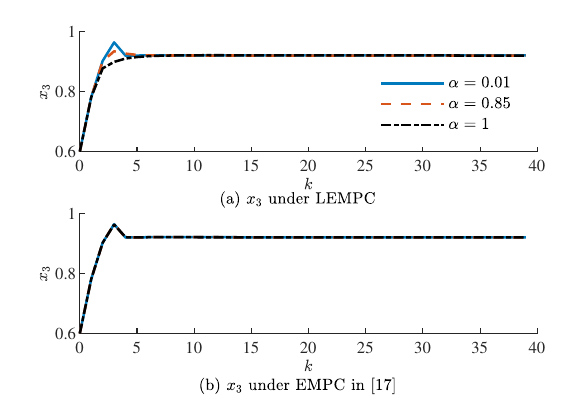}
	\caption{State $x_3$ trajectories under LEMPC and EMPC in \cite{defeng2019input}}
	\label{FIG:7}
\end{figure}

\begin{figure}  
	\centering
	\includegraphics[width=.5\textwidth]{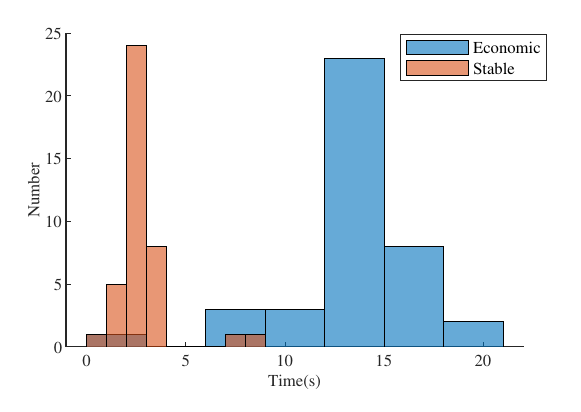}
	\caption{Statistical results on the optimization times}
	\label{FIG:8}
\end{figure}

To demonstrate the control advantages of our LEMPC, we choose the EMPC in \citet{defeng2019input} as a comparison. Both algorithms are based on our CKLR predictor for fairness. Table 4 lists the average economic performance over the transient process with initial state [0.9, 0.82, 0.6] and different $\alpha  = 0.01,{\rm{ }}0.85$ and 1.0. Under the same parameter $\alpha$, the EMPC in \citet{defeng2019input} has better economic performance. However, this performance deterioration of LEMPC can be compensated by decreasing $\alpha$, and the performance difference gradually disappears when $\alpha$ tends to 0. Fig.7 further pictures the corresponding state trajectories. It can be observed that increasing $\alpha$ accelerates the convergence of the states in the price of economic performance, and LEMPC significantly improves the convergence speed. In particular, we emphasize that the LEMPC recovers the control effect of stabilized MPC for $\alpha=1$, but EMPC in \citet{defeng2019input} fails to accelerate convergence even under $\alpha=1$. Thus, compared to EMPC in \citet{defeng2019input}, LEMPC is more flexible regarding the trade-off between performance and stabilization.

Fig.8 provides the statistical results on the optimization times of 40 simulations to illustrate the real-time capabilities of LEMPC. The maximum sum of the computational time is 23.8887 s, less than $T_s = 30$ s. Economic optimization takes a longer time due to extra constraints (32f). Although the optimization burden may be reduced via PKI and SPKI schemes, the errors are magnificent and exceed the error bound 0.00881 in Fig.5. To summarize, the proposed LEMPC scheme is reliable for the CSTR and exhibits well synthetic performance. 

\section{Conslusion}

A novel CKLR nonparametric regression method and an elegant learning EMPC scheme for constrained, uncertain nonlinear systems with unknown dynamics were proposed. Using the clustering and kernel techniques, the CKLR algorithm significantly improved the prediction accuracy and reduced the online prediction time. Moreover, based on the stabilized MPC value function, we constructed the Lyapunov-like contraction constraint for EMPC under two learning error analyses, which ensured the ISS and recursive feasibility of the closed-loop system. In particular, the proposed scheme allowed the closed-loop system to compromise convergence speed and performance as well as recover the stabilized MPC scheme under a specific parameter design. Two simulation experiments verified the merits of the proposed CKLR and LEMPC scheme. 

Future work will focus on developing an online learning-based EMPC scheme with theoretical properties and designing a fast LEMPC by introducing a variable-horizon mechanism. In addition, the proposed LEMPC algorithm can be implemented in actual plants, such as vehicles and chemical processes, which possess plentiful historical data for control purposes.

\bibliographystyle{unsrtnat}
\bibliography{references}  %%% Uncomment this line and comment out the ``thebibliography'' section below to use the external .bib file (using bibtex) .




% 	\bibitem{keshet2016prediction}
% 	Keshet, Renato, Alina Maor, and George Kour.
% 	\newblock Prediction-Based, Prioritized Market-Share Insight Extraction.
% 	\newblock In {\em Advanced Data Mining and Applications (ADMA), 2016 12th International 
%                       Conference of}, pages 81--94,2016.

% \end{thebibliography}

\end{document}